\pdfoutput=1
\documentclass[11pt]{article}
\usepackage[margin=1in]{geometry}
\usepackage[T1]{fontenc}
\usepackage{lmodern}
\usepackage{microtype}
\usepackage{graphicx}
\usepackage{booktabs}
\usepackage{array}
\usepackage{amsmath}
\usepackage{xcolor}
\definecolor{pickcolor}{HTML}{008A3E}
\definecolor{changecolor}{HTML}{B4236E}
\newcommand{\pickpart}[1]{\textcolor{pickcolor}{#1}}
\newcommand{\changepart}[1]{\textcolor{changecolor}{#1}}
\usepackage[round,authoryear]{natbib}
\usepackage{xurl}
\usepackage[hidelinks]{hyperref}
\usepackage{caption}
\usepackage{placeins}
\newcommand{\pp}{\,pp}

\title{When the Rule-Maker Runs the World Championship: Late Patches and Procedural Accountability in \emph{League of Legends}}
\author{Haewoon Kwak\\Indiana University Bloomington\\\texttt{hwkwak@iu.edu}}
\date{}

\begin{document}
\maketitle

\begin{abstract}
Riot Games, which runs the \emph{League of Legends} World Championship (Worlds), also changes the game shortly before each tournament. This concentration of authority raises a governance question: Who scrutinizes how the benefits and burdens of late rule changes are distributed? We examine all ten championships from 2016 to 2025, combining $1{,}097$ games, $215$ team-years, and LLM coding of $1{,}090$ champion changes across $40$ patches. On average, $33\%$ of a team's prior picks were on champions with net nerfs in the patches it had not played before Worlds, and $8\%$ were on champions with net buffs. Within a year, the median gap between the most and least affected teams was $23$ percentage points for nerfed picks and $12$ for buffed picks. Nerfs fell mainly on the most-contested champions and buffs on less-contested ones, as Riot's published balancing criteria anticipate. Pools concentrated on the prevailing meta therefore had more negative net exposure, and even the winners of a year's four major regional leagues differed in net exposure about as much as four randomly drawn teams. Higher buff exposure was associated with additional opponent bans on buffed champions, although its size depends on the model. Among quarterfinalists, the most buff-exposed team of each year reached six of ten finals and won four titles. The team with the most negative net exposure reached two finals and won one. These exploratory descriptions identify unequal exposure, not causal gains or losses. The association between net exposure and winning remains uncertain, at $+1.3$ percentage points per standard deviation with a $95\%$ interval from $-0.9$ to $+3.7$. We argue that scrutiny of late rulemaking should not depend on proving that a patch changed who won Worlds. Earlier patch commitments, team-level impact disclosure, independent review, and a formal opportunity to challenge decisions would make publisher authority more accountable.
\end{abstract}

\section{Introduction}

Who should be able to change the conditions of a sports competition just before it begins? Established sports offer procedures for making such decisions. In football, changes to the Laws of the Game pass through the International Football Association Board (IFAB), where FIFA and the four British associations share voting authority and a three-quarters majority is required \citep{ifaborganisation}. The 2025/26 changes were approved in March for a normal effective date of July 1, with provisions for competitions already under way \citep{ifab2025circular}. Football's rules also evolve, and its institutions combine regulatory and commercial roles. The relevant comparison is how rule changes are scheduled, justified, and subjected to collective scrutiny.

Compared with football, publisher-run esports concentrate this authority to an unusual degree, because the rule-maker also runs the championship. For example, Riot Games owns \emph{League of Legends} (LoL), develops its balance patches, and runs its World Championship (``Worlds'') \citep{taylor2012,scholz2019,karhulahti2017}. Patches strengthen or weaken champions, items, and systems, changing which resources teams can bring to competition. Even at Worlds, the designated patch arrives close to the event. Across the ten Worlds we study, its release preceded the start of the championship by $5$ to $21$ days (Appendix~\ref{app:timing}). Riot explicitly uses this discretion to shape the competition. Its 2024 Worlds patch notes describe changes intended to produce a ``varied and action-packed Worlds,'' including interventions in champion selection and lane-swap incentives \citep{riot1418}. The company making these choices also has a commercial interest in the event and the game \citep{karhulahti2017,holden2017}.

Riot releases patch notes that explain changes along with selected design goals. It has also published criteria for balancing around professional play \citep{riot2020proplay}. These documents help teams and fans follow what changes, but by themselves they do not make balancing decisions accountable. Criteria defined over champions' aggregate presence, and a list of the resulting changes, cannot establish how competing objectives were weighed in a given patch, which teams' established pools bear the costs, or how affected participants can obtain independent review. The governance concern is therefore present even when balancing decisions are made in good faith. A common patch can have unequal consequences for teams that qualified using different champion pools and often different versions of the game across regions. Those consequences deserve scrutiny before the championship begins, including for teams that receive little attention after early elimination.

We investigate all ten World Championships from 2016 to 2025 to examine how these consequences are distributed and the limits of judging them from match results. Our central argument is \emph{procedural}. The authority to revise a championship's competitive conditions shortly before play should carry obligations to disclose team-level incidence, justify tradeoffs, and provide a meaningful route to challenge the decision. The empirical analysis makes part of that incidence observable. It does not recover internal publisher deliberations or identify how a tournament would have unfolded on another patch. We measure a team's exposure from its prior official picks and the coded changes in patches it had not played before Worlds. This measure does not capture lost champion mastery, the practice time needed to adapt, or changes in title probability.

We make four contributions to this argument.
\begin{enumerate}
\item \textbf{Who is affected.} Using team-specific patch windows and prior pick shares, we show that nerfs concentrate on the most-contested champions and buffs on less-contested ones, as Riot's published presence-based criteria anticipate. Teams specialized in the prevailing meta therefore carry more negative exposure. Even among the winners of each year's four major regional leagues, exposure differs about as much as among randomly drawn teams (Sections~\ref{sec:data} and~\ref{sec:pool}).
\item \textbf{How exposure may reach competition.} Greater buff exposure is associated with additional opponent bans on a team's buffed champions. The size of this association depends on how pick share is modeled (Section~\ref{sec:draft}).
\item \textbf{What results can and cannot settle.} Associations between exposure and winning remain uncertain in ten championships, and our outcome analysis therefore does \emph{not} settle disputes over late patches in either direction. Cases on both sides, the most buff-exposed and the most negatively exposed quarterfinalists, illustrate why a dispute about a particular title is hard to settle from match results at all. Each title is a single event without an observable counterfactual (Sections~\ref{sec:results} and~\ref{sec:cases}).
\item \textbf{Procedural accountability.} We propose safeguards that operate before competition and argue that they are justified on their own, without proof that a patch changed who won Worlds (Section~\ref{sec:discussion}).
\end{enumerate}

\section{Background and related work}
\label{sec:background}

\paragraph{Patches and the World Championship.} LoL is a popular multiplayer online battle arena (MOBA) game in which two teams of five players try to destroy the opposing team's base, with each player controlling a character called a champion. Teams ban and pick champions in a draft. Each team can ban five champions per game, three in 2016, and picks five, and a champion picked or banned by one team is unavailable to the other in that game. In 2025, fearless draft additionally made champions picked by either team in earlier games of the series unavailable \citep{riot2025fearless}. Riot changes champion statistics and abilities through patches. Changes that strengthen a champion are called buffs, and changes that weaken it nerfs.

In 2020, Riot published criteria for balancing around professional play \citep{riot2020proplay}. At the Mid-Season Invitational (MSI) and Worlds, a champion receives a nerf when its pick-and-ban presence, the share of games in which it is picked or banned, reaches $90\%$ in one patch or $80\%$ in two consecutive patches. Champions below $5\%$ presence are considered underpowered at the professional level, and diversity goals guide buffs, such as at least ten picks per position with $5\%$ presence or more for Worlds 2020. Low professional presence alone, however, does not guarantee a buff. Large or systemic changes are to be minimized in the lead-up to Worlds. Riot's balance framework treats professional play as one of four player groups. It flags a champion for nerfs if it overperforms in any group and for buffs if it underperforms in all of them, and it leaves room for interventions on game-health grounds \citep{riot2020framework}. These criteria concern champions' aggregate presence and performance, not how changes fall on the teams that qualified with particular pools. Worlds uses a designated patch, while participants' last officially played patches can differ.

Kica et al.\ study patches and champion win rates \citep{kica2016}. He et al.\ examine patch-related performance changes in LoL, including variation across players and team compositions \citep{he2021}. We build on this work by measuring team-specific exposure to coded champion changes around Worlds and by distinguishing exposure, draft associations, and outcome uncertainty.

\paragraph{Rule changes and competitive balance.} Sports economics treats the distribution of advantage and uncertainty of outcome as central features of a contest \citep{rottenberg1956,neale1964,fort1995,szymanski2003}. Rule changes can alter incentives and competitive balance \citep{moschini2010,judde2013}. Equipment changes can also affect performance, as in baseball's investigation of home-run rates \citep{albert2018}. Worlds brings together frequent software revision, a short interval before championship play, and control of both game design and event operation within the publisher. This combination makes the process for authorizing changes a substantive part of competitive governance.

\paragraph{Esports governance and integrity.} Work on esports governance emphasizes publishers' ownership of the intellectual property underlying professional competition \citep{karhulahti2017,holden2017}. Sport integrity is a broader, contested concept \citep{gardiner2017}. We connect this institutional concern to the distribution of specific pre-event changes across teams. The contribution is an empirical basis for scrutiny of publisher discretion, with procedural proposals that address the concern. We do not measure audience trust, participant perceptions, or the effectiveness of existing consultation. Nor do we assume that esports lacks all independent dispute resolution. Riot's EMEA mechanism refers contractual and financial disputes between teams, players, and coaches to independent arbitrators \citep{riot2024disputes}. We argue that similar procedural mechanisms are essential for reviewing championship patch decisions.

\section{Data and measures}
\label{sec:data}

\subsection{Matches}
We use Oracle's Elixir match data \citep{oracleselixir}, including game dates, patches, picks, bans, and results. Oracle's Elixir lacks some games that Worlds participant teams played before Worlds, mostly in 2016--2019, such as Rift Rivals and the 2016 International Wildcard Qualifier. We restore $547$ games of that year's participants in the $365$ days before Worlds from Leaguepedia's scoreboards \citep{leaguepedia}, matching the two sources by champion picks, timing, and linked team identities (Appendix~\ref{app:supplement}). On the $15{,}793$ participants' games matched in both sources, they agree on the winner in $99.99\%$ and on the bans in $99.7\%$. We also correct Oracle's Elixir records that fail consistency checks. Records that copy another game are excluded, and sides recorded under the wrong team are moved to the team whose players played them. A missing winner is set only when at least two sources agree and none contradicts. These sources are the end-of-game statistics in Oracle's Elixir, Leaguepedia, a Games of Legends game page \citeyearpar{golgg}, and a Liquipedia series score \citeyearpar{liquipedia}. Games whose Oracle's Elixir and Leaguepedia winners differ are excluded. Table~\ref{tab:corrections} reports the corrections, none of which touches a Worlds game. We exclude three MSI 2022 games that Riot voided and replayed because of a latency error and keep the replays \citep{riot2022latency}.

We study 2016--2025 because earlier years have too few regional games in the source for pre-tournament measures. Each Worlds is defined by its designated patch and dates, covering play-ins and the main event. The 2023 Worlds Qualifying Series on October 9 decided the last participating team before play-ins began on October 10. Its games therefore count as pre-event history, not as Worlds games \citep{riot2023wqs,riot2023dates}. Our final dataset contains $1{,}097$ games and $215$ team-years. All $215$ team-years have at least ten official games in the preceding $100$ days, and therefore the main exposure--outcome model includes all $1{,}097$ games. Knockouts are the last seven series of each event. We call the teams that won the last domestic playoff final before Worlds in the LCK, LPL, LEC (EU LCS before 2019), and LCS (LTA in 2025) regional winners.

\subsection{Patch window and blind coding}
\label{sec:coding}
For each year we collected the official notes of every patch after the earliest last-regular-season patch among the participants, up to and including the Worlds patch (hereafter, the coded patches). The $40$ patches contain $1{,}093$ champion-level changes. Three applied only to the All Random, All Mid (ARAM) mode, leaving $1{,}090$ changes on Summoner's Rift, the map used in professional play. Each change was coded for direction (buff, nerf, adjustment, rework, or bug fix) and for magnitude from $1$ to $3$. The signed score is the magnitude for buffs, minus the magnitude for nerfs, and zero otherwise.

We used Claude Opus 5.5 (Anthropic) agents for the coding, following work on large language model (LLM) text annotation \citep{gilardi2023,ziems2024}, with the additional checks described below. The agents were instructed to work from the patch-note text only and not to look up esports results, winners, or professional picks, and an audit of their tool calls found no access to match data (Appendix~\ref{app:coding}).

We check the codes in three ways. First, the author coded a stratified sample of $120$ changes with the same coding scheme, without seeing the LLM codes. Direction agreed in $87.5\%$ and the sign in $89.2\%$ (Cohen's $\kappa = 0.84$), and most disagreements concerned whether a change that mixes increases and decreases is an adjustment. Second, a separately run coder using the same model re-coded $309$ champion--patch pairs from three years and agreed on the sign in $99.4\%$ of them ($\kappa = 0.99$). Third, independently of the patch notes' wording, we compared each champion's values in the game data with those of the preceding patch. Where the values moved in only one direction, the coded sign matched it in $94.9\%$ of $450$ changes, and in $98.4\%$ of the $434$ that the LLM coded as buffs or nerfs. The codes are also consistent with professional behavior. Among champions selectable at that Worlds, those coded as buffed rose in pick-and-ban presence at Worlds relative to the preceding regional games by $7.8$\pp{} on average, and nerfed champions fell by $12.6$\pp{} (correlation between coded score and presence change $r = 0.42$, $n = 504$). Additionally, we reran the full pipeline after recoding the changes that contradict the game data, ignoring magnitudes, or taking directions from the game data where they are comparable. The direction and statistical conclusions of the main results remained unchanged (Appendix~\ref{app:coding}).

Item, rune, and system changes were also coded, but independent coders agreed on their direction in only $52$--$71\%$ of cases, because which champions such a change helps or hurts must often be inferred. Exposure is therefore restricted to direct champion changes. It is a partial measure and not a lower bound on the total effect, because omitted changes could reinforce or offset champion changes.

\subsection{Exposure}
We define a team's exposure to a Worlds patch from two quantities: \pickpart{how often the team played each champion} and \changepart{how much that champion was buffed or nerfed in the patches the team had not yet played}. We use \pickpart{$w_i(r,c)$} to represent the share of team $i$'s games in the preceding $100$ days in which it played champion $c$ in role $r$. Likewise, we use \changepart{$\delta_i(c)$} to represent champion $c$'s summed signed score over the coded patches newer than the version of team $i$'s last official game before Worlds, through the Worlds patch. This version-based definition differs from one based on release dates, because official matches are often played on an older patch after a newer one has been released. Then, the team's net exposure can be formally written as:
\begin{equation}
\text{net exposure}_i = \sum_{r,c} \pickpart{w_i(r,c)}\,\changepart{\delta_i(c)}.
\label{eq:exposure}
\end{equation}
By definition, a team that had already played the Worlds patch before Worlds began has zero net exposure, because \changepart{$\delta_i(c)$} then sums over no patch versions. Keeping roles separate lets exposure be split by role (Section~\ref{sec:pool} and Appendix~\ref{app:roles}). Because net exposure combines buffs and nerfs into one number, we also define buff exposure and nerf exposure to see how each acts separately. They use the positive and negative parts of each champion's net change:
\begin{align}
\text{buff exposure}_i &= \sum_{r,c} \pickpart{w_i(r,c)}\,\max(\changepart{\delta_i(c)},0), \label{eq:buff}\\
\text{nerf exposure}_i &= \sum_{r,c} \pickpart{w_i(r,c)}\,\max(-\changepart{\delta_i(c)},0). \label{eq:nerf}
\end{align}
Both are nonnegative, and each grows with the size of the corresponding buffs or nerfs. Net exposure is therefore buff exposure minus nerf exposure.

For a simpler description that ignores the size of the change, we also report the percentage of a team's picks on champions with a net buff or a net nerf:
\begin{align}
\text{buffed pick share}_i &= \frac{1}{5}\sum_{r,c:\,\changepart{\delta_i(c)}>0} \pickpart{w_i(r,c)}, \label{eq:buffshare}\\
\text{nerfed pick share}_i &= \frac{1}{5}\sum_{r,c:\,\changepart{\delta_i(c)}<0} \pickpart{w_i(r,c)}. \label{eq:nerfshare}
\end{align}
The factor $1/5$ normalizes the sum to a share of all picks, because a team plays five champions in each game.

Exposure scores are standardized within year over eligible teams, and pick shares are reported as percentages. The same $\delta_i(c)$ enters the pool, ban, and pick analyses. The number of coded patches in a team's window ranges from zero to six (mean $3.2$).

\subsection{Pre-tournament strength}
To adjust outcome associations for observed pre-tournament strength, we rate every team with a Bradley--Terry model \citep{bradley1952} fitted to all games in the $365$ days before each Worlds. The model has a blue-side term, league effects linked through international events, penalized team deviations, and exponential down-weighting of older games with a half-life of $240$ days. It correctly classifies the winner in $68.7\%$ of Worlds games with a log loss of $0.608$, or $0.611$ when its two hyper-parameters, the half-life and the penalty on team deviations, are chosen leave-one-year-out. Aggregating its game forecasts over the games actually played, which conditions on each series' realized length and sides, classifies $72.3\%$ of series correctly. It underrates LCK teams by $5.2$\pp{} per cross-region game. Appendix~\ref{app:strength} reports sensitivity to region controls and to the team-rating window.

\subsection{Inference}
We model match results with game-level logistic regressions. The outcome is whether the blue side won, and the predictors are the two teams' differences in pre-event rating and in exposure. For readability, we report each coefficient $\beta$ as $100\beta/4$. This is the change in win probability, in percentage points, for a one-standard-deviation (SD) increase in a team's exposure relative to its opponent's in an evenly matched game. Because games at the same Worlds are not independent, the main $95\%$ interval resamples whole events $2{,}000$ times. With only ten events to resample, the interval may contain the true value less often than its nominal $95\%$.

As a supplementary test, we shuffle exposure among the teams of the same year $2{,}000$ times and refit the model each time. The test statistic is the coefficient divided by its model-based standard error, and the two-sided $p$-value is $(1+\#\{|z^{*}|\geq|z|\})/(1+B)$, where $z$ is the observed statistic, $z^{*}$ a shuffled one, and $B = 2{,}000$ \citep{efron1993,good2005}. Shuffling treats exposure as interchangeable among the teams of a year when it has no association with winning. Dividing by the standard error keeps the test closer to its nominal error rate than the raw coefficient when exposure is not fully interchangeable \citep{chung2013}. We call this the studentized permutation test. It still does not guarantee validity for observational exposures. Section~\ref{sec:results} therefore checks the test on outcomes simulated from a rating-only model, with no exposure effect, a check that holds only under that model.

To study bans, we define a team's core champions as those it picked in at least $10\%$ of its games in the same $100$ days, that is, champions with $\sum_r \pickpart{w_i(r,c)} \geq 0.1$. For each Worlds game, we count the opponent's bans on the team's core champions and compare this count with an expected count, based on how often those champions were banned in the other games of the same Worlds. Excess bans are then the observed minus the expected count. Champions that could not be selected are left out of both counts. Riot disabled ten champions for all or part of a Worlds. Eight were new or reworked champions released too close to the event, and two were disabled for bugs. They are listed with sources in the replication files. In 2025, fearless draft also made champions picked earlier in the series unavailable.

Two measures describe the style of a team's champion pool, both from the same $100$ days. Pool breadth averages, over the five roles, the effective number of champions the team played in each role, the exponential of the Shannon entropy of its pick shares \citep{hill1973,jost2006}:
\begin{equation}
\text{breadth}_i = \frac{1}{5}\sum_r \exp\Big(-\sum_c \pickpart{w_i(r,c)}\ln \pickpart{w_i(r,c)}\Big).
\label{eq:breadth}
\end{equation}
The effective number is the number of equally played champions that would give the same entropy. 

Meta concentration weights each champion by its pick-and-ban presence $\pi_{-i}(c)$ in other teams' games of the same $100$ days:
\begin{equation}
\text{meta concentration}_i = \frac{1}{5}\sum_{r,c} \pickpart{w_i(r,c)}\,\pi_{-i}(c).
\label{eq:meta}
\end{equation}
A team with high meta concentration relies on champions that other teams also often pick or ban. Both measures are standardized within year.

Teams are linked across documented renames and whole-team acquisitions, ten lineages listed with sources in Appendix~\ref{app:orgs}. Regressions of exposure on breadth and meta concentration use standard errors clustered by organization, with Worlds-level alternatives in Appendix~\ref{app:presence}, and the champion-level ban model clusters by team-year (Appendix~\ref{app:bans}). Champion comparisons and secondary models are exploratory, and reported $p$-values are not corrected for the number of analyses.

For the descriptive cases in Section~\ref{sec:cases}, we rank buff exposure and net exposure among quarterfinalists and, as checks, among all participating teams and by nerf exposure. All such comparisons are post hoc, and their tournament finishes are descriptive.
\section{Results}

\begin{figure}[tbp]
\centering
\includegraphics[width=\textwidth]{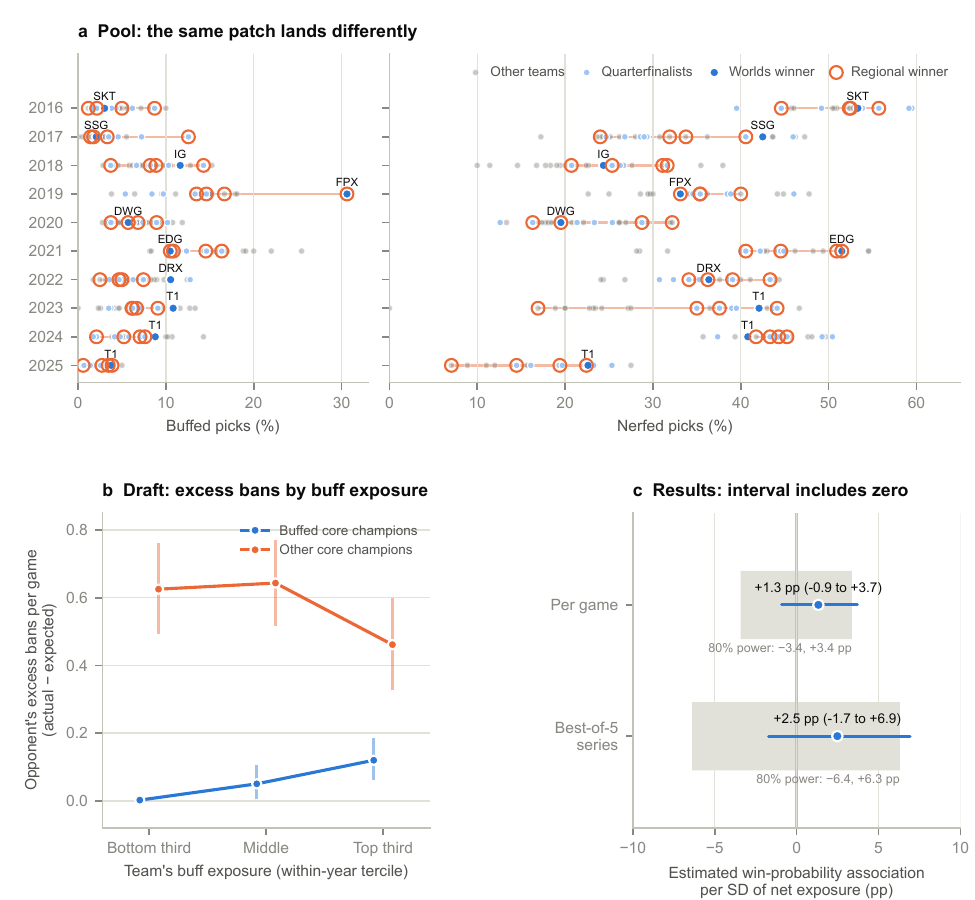}
\caption{Late patches at three levels, 2016--2025. \textbf{a}, Each team's pre-Worlds picks on champions with net buffs (left) and nerfs (right) in the patches it had not played before Worlds. \textbf{b}, Excess opponent bans on core champions by buff-exposure tercile. \textbf{c}, Association of net exposure with win probability per SD, with $80\%$ power thresholds in gray.}
\label{fig:levels}
\end{figure}

Figure~\ref{fig:levels} summarizes the results at three levels: team pools, drafts, and match results. We first show which teams the late patches affected most and why nerfs concentrate on the contested meta (Section~\ref{sec:pool}). We then ask whether buff exposure reached the draft through opponent bans (Section~\ref{sec:draft}) and what match results can and cannot establish about exposure and winning (Section~\ref{sec:results}). Finally, we describe the most visible cases, the most buff-exposed and the most negatively exposed quarterfinalists of each year (Section~\ref{sec:cases}).

\subsection{Who pays for balance? Nerfs follow the contested meta}
\label{sec:pool}
The patches a team had not played before Worlds changed many of the champions it relied on. On average, $33\%$ of a team's pre-Worlds picks were on champions with net nerfs in those patches and $8\%$ on champions with net buffs (Figure~\ref{fig:levels}a). Within a single year, the median gap between the most and least affected teams was $23$ percentage points in nerfed picks and $12$ in buffed picks. The burden also differs by role. On average, $46\%$ of bottom-lane and $41\%$ of jungle picks were on champions with net nerfs, compared with $24\%$ in top lane and $22\%$ in support. In $133$ of the $201$ team-years with a unique largest role, jungle or bottom lane had the largest absolute net exposure (Appendix~\ref{app:roles}). Exposure varies even among each year's most successful regional teams. Within a year, the four regional winners differed by a median of $1.80$ SD in net exposure, close to the $2.06$ SD expected for four randomly drawn teams, whose median range has a $90\%$ interval of $1.55$--$2.61$. In 2023, the regional winners' nerfed-pick shares ranged from $17\%$ (G2 Esports) to $44\%$ (NRG).

\begin{figure}[tbp]
\centering
\includegraphics[width=\textwidth]{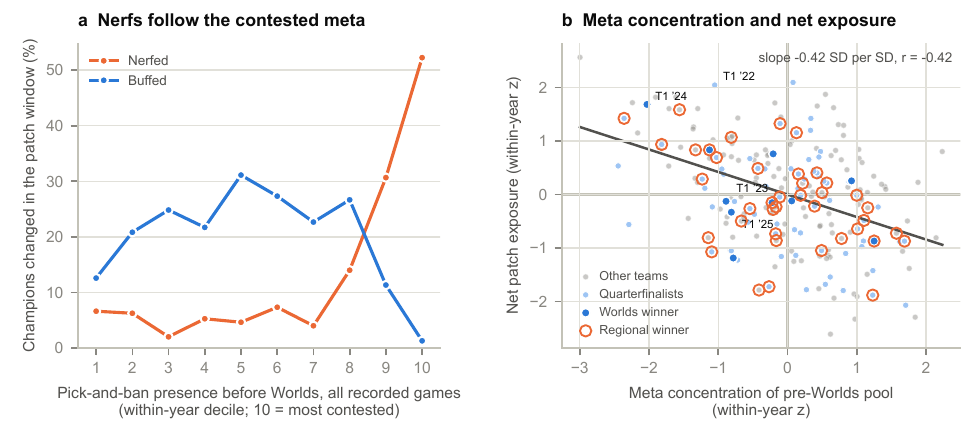}
\caption{Where nerfs and buffs fall. \textbf{a}, Share of champions with a net nerf or buff across each year's coded patches, by pre-Worlds pick-and-ban presence decile; nerfs concentrate on the most-contested champions. \textbf{b}, Net exposure against meta concentration for $215$ team-years, both standardized within year.}
\label{fig:mechanism}
\end{figure}

Nerfs concentrate on the most-contested champions, those with the highest pick-and-ban presence, and buffs on less-contested ones. Figure~\ref{fig:mechanism}a sorts champions by their pick-and-ban presence in the $100$ days before Worlds, counting with equal weight every game in the match data, from all leagues Oracle's Elixir covers, including development leagues, plus the restored games. For this comparison we sum changes across each year's coded patches, whereas team exposures use each team's own unplayed versions.

Figure~\ref{fig:mechanism}a covers $1{,}502$ champion-years with nonzero presence. In the most-contested decile, $52\%$ of champions were nerfed and $1\%$ buffed. In the bottom eight deciles, $6\%$ were nerfed and $23\%$ buffed. The pattern holds when presence counts only the four major regional leagues or is measured in the $100$ days before each year's first coded patch was released (Appendix~\ref{app:presence}). Riot's published criteria anticipate this pattern, since high professional presence triggers nerfs and buffs aim to widen the set of viable picks \citep{riot2020proplay}. The criteria are defined over champions, however, and do not say how the resulting changes fall on teams whose pools differ.

Teams whose champion pools concentrate on the contested meta therefore experience more negative exposure. A one-SD increase in meta concentration predicts $+0.35$ SD of nerf exposure ($p < 0.001$), $-0.20$ SD of buff exposure ($p = 0.004$), and $-0.40$ SD of net exposure ($p < 0.001$; Figure~\ref{fig:mechanism}b). Interestingly, pool breadth adds no detectable association once meta concentration is included. The line in Figure~\ref{fig:mechanism}b is an unadjusted fit, while these coefficients adjust for breadth. The associations of meta concentration with nerf exposure and with net exposure also hold when dependence among teams at the same Worlds is allowed for ($p \leq 0.01$) and under every presence definition. Its association with buff exposure does not (Appendix~\ref{app:presence}), and we draw no conclusion from it. Among the $40$ regional winners alone, meta concentration is also negatively correlated with net exposure, with meta concentration spanning $-2.4$ to $+1.7$ SD ($r = -0.50$, against $r = -0.42$ for all eligible teams).

A linear model of net exposure on pool breadth and meta concentration, fitted on the other nine years, ranks the held-out year's teams with an average Spearman correlation of $0.35$ (range $-0.31$ to $0.61$). In other words, pool style alone ordered each held-out year's teams by net exposure in the right direction in nine of the ten years, though only moderately. The exception, 2020, is the year in which the late patches nerfed the fewest of the most-contested champions ($13\%$, against $35$--$79\%$ in other years). Averaged over the other nine years, the correlation is $0.42$.

\subsection{Do teams get to use their buffs? Bans follow buff exposure}
\label{sec:draft}
A buff helps a team only if it can pick the buffed champion. Opponents can take that option away by banning the champion in the draft. We ask whether they did so more often against teams with high buff exposure.

Teams in the top third of buff exposure faced $0.22$ opponent bans per game on buffed core champions against $0.10$ expected, an excess of $0.12$, where the expected count comes from how often the same champions were banned in the other games of the same Worlds. The bottom third's excess was approximately zero (Figure~\ref{fig:levels}b, where bars are $95\%$ bootstrap intervals over teams).

More bans on buffed champions could mean that these champions were banned often against everyone or that opponents banned them more against this team. Across the $212$ team-years with nonzero buff exposure, buff exposure correlates with the fraction of that exposure removed by opponent bans ($r = 0.22$, $p < 0.001$). Splitting that fraction into the part expected from the other games of the same Worlds and the excess beyond it, the correlation comes from the excess ($r = 0.24$, $p < 0.001$), not the expected part ($r = 0.03$, $p = 0.59$). Opponents thus banned these champions more against high-buff-exposure teams than against other teams at the same Worlds.

Bans on a team's other core champions moved the other way. Across the thirds of buff exposure, excess bans on other core champions decreased from $0.63$ to $0.46$. Within a team-year, an additional excess ban on a buffed champion was associated with $0.36$ fewer on other core champions ($p < 0.001$), consistent with each team's fixed number of bans per game.

In the champion-level model, an opponent's ban probability rises with the team's prior pick share of a champion. A one-SD increase in team buff exposure makes this rise steeper by $11.8$\pp{} per unit of pick share for buffed champions ($p = 0.004$) and flatter by $2.5$\pp{} for other champions ($p = 0.055$). For two buffed champions whose pick shares differ by $10$ points, the gap in ban probability therefore widens by about $1.2$\pp{} ($0.1 \times 11.8$\pp{}) per SD of buff exposure. The steeper rise for buffed champions is well supported, whereas the evidence that the team's other champions were banned less is weak. Restricted to series game 1, the rise for buffed champions is steeper by $13.1$\pp{} ($p = 0.012$). The association therefore does not require adaptation to earlier games of the same series, although teams may still learn from other series. Because team buff exposure contains each champion's own pick share, how much steeper the rise is depends on how pick share is modeled. Across three alternative specifications it stays steeper, by $9.6$ to $13.3$\pp{} (Appendix~\ref{app:bans}).

\subsection{Does exposure change who wins? The evidence is inconclusive}
\label{sec:results}
Match results show no clear association between exposure and winning. Net exposure is associated with $+1.3$\pp{} in win probability per SD, with a $95\%$ event-bootstrap interval from $-0.9$ to $+3.7$ (studentized permutation $p = 0.333$, $n = 1{,}097$; Figure~\ref{fig:levels}c). In a model containing both buff and nerf exposure, the buff coefficient is $+2.2$\pp{} (interval $-1.2$ to $+5.2$, $p = 0.111$), but its size depends on how change directions are coded (Appendix~\ref{app:coding}). Replacing the pre-tournament rating with other strength controls, or dropping it, does not change this picture (Appendix~\ref{app:strength}). Figure~\ref{fig:levels}c also converts the per-game estimate to a best-of-five series, an illustrative conversion that assumes independent games with constant win probability (Appendix~\ref{app:power}).

Exposure could also reach match results through the draft, where opponents ban a team's champions and the team picks its own. We therefore also relate two draft measures to winning. These are excess bans and pick advantage, the net coded change of a team's five picked champions minus that of its opponent's picks. They have associations of $-1.2$ and $-1.5$\pp{} per SD, respectively, with intervals including zero.

To judge what associations the $1{,}097$ Worlds games can detect, we simulated outcomes from a rating-only model, holding the observed match schedules and exposures fixed. The studentized test rejected $5.1\%$ of $2{,}000$ null simulations. Estimated power was $49\%$ at a positive $2.5$\pp{} slope and $83.2\%$ at $3.5$\pp{}, with an interpolated $80\%$ threshold of about $3.4$\pp{} in either direction (Appendix~\ref{app:power}). Under the rating model, the test thus rejects about as often as it should when exposure has no association with winning, but these games would usually reveal only per-game associations of about $3.4$\pp{} per SD or larger. Our outcome analysis establishes neither a causal advantage from late patches nor its absence, because the observed interval includes zero.

\subsection{Which teams stand out? The most buff-exposed and most negatively exposed quarterfinalists}
\label{sec:cases}
We illustrate the exposure distribution with prominent cases at both extremes. Both are drawn from quarterfinalists, the most visible teams at each Worlds. This frame, however, keeps only teams that advanced after the patch and drops exposed teams that exited early. We therefore also report counts over all participating teams and check whether a ranking by nerf exposure picks the same teams.

\paragraph{The most buff-exposed quarterfinalist.}
Table~\ref{tab:champions} lists the quarterfinalist with the highest buff exposure each year. That team reached six of ten finals and won four titles. From 2022 to 2025 it was T1 each year, and T1 reached four finals and won three. These counts are descriptive. We do not test them against a random-quarterfinalist benchmark, which would ignore team strength, bracket position, and recurring organizations.

The corresponding highest-net-exposure quarterfinalist won twice. Across all participating teams, the most buff-exposed team reached two finals and won one title, and six of the ten exited before the quarterfinals. T1's pools were the broadest among quarterfinalists in 2022--2025, and their ranks from least to most meta-concentrated were $2$, $1$, $1$, and $3$. None of T1's buff exposure came from the most-contested decile of pre-Worlds presence (Section~\ref{sec:pool}). Pooled across these four years and weighted by raw buff exposure, $64\%$ came from champions with at least $20\%$ presence in all recorded games during the $100$ days before the year's first coded patch was released. This share ranged from $17\%$ to $79\%$ across years. In each year, $65$--$94\%$ came from champions also net-buffed in at least one other quarterfinalist's pool, using each team's own unplayed versions. For example, in 2022, Lee Sin made up $28\%$ of T1's jungle picks and $65\%$ of its buff exposure across roles. After a descriptive adjustment for breadth and meta concentration with the regression of Section~\ref{sec:pool}, T1 ranked first among quarterfinalists in 2022 and 2025, third in 2023, and fourth in 2024. Although these data locate where T1's buff exposure came from, they show neither why particular champions were buffed nor any preferential treatment, and we do not test for such treatment. The T1 case thus makes differential exposure visible around high-profile outcomes but does not identify an effect on who won Worlds. More broadly, whether teams differ systematically in buff exposure cannot be settled from these data. The answer depends on whether renamed teams count as one organization and rests largely on a single lineage (Appendix~\ref{app:orgs}).

\begin{table}[htbp]
\centering
\small
\caption{The quarterfinalist with the highest buff exposure each year.}
\label{tab:champions}
\begin{tabular}{lllc}
\toprule
Year & Most buff-exposed quarterfinalist & Finish & Worlds winner \\
\midrule
2016 & EDward Gaming & Quarterfinal & SK Telecom T1 \\
2017 & Longzhu Gaming & Quarterfinal & Samsung Galaxy \\
2018 & Fnatic & Runner-up & Invictus Gaming \\
2019 & FunPlus Phoenix & \textbf{Winner} & FunPlus Phoenix \\
2020 & DRX & Quarterfinal & DAMWON Gaming \\
2021 & MAD Lions & Quarterfinal & EDward Gaming \\
2022 & T1 & Runner-up & DRX \\
2023 & T1 & \textbf{Winner} & T1 \\
2024 & T1 & \textbf{Winner} & T1 \\
2025 & T1 & \textbf{Winner} & T1 \\
\bottomrule
\end{tabular}
\end{table}

\paragraph{The most negative net exposure among quarterfinalists.}
Focusing on the most buff-exposed team can overlook teams whose frequently picked champions were nerfed. We therefore select, for each year, the quarterfinalist with the lowest net exposure (Table~\ref{tab:burden}). The table also gives the number of coded patches newer than each team's last officially played version. A negative net exposure means that, after combining buffs and nerfs, the coded changes weakened the champions the team had been playing.

Six of these ten teams lost in the quarterfinals, two in the semifinals, and one reached the final and lost it. One, Samsung Galaxy in 2017, won the title. The most negative exposure among quarterfinalists thus did not preclude a championship, just as high buff exposure did not guarantee one. The recent cases were Weibo Gaming in 2023 ($-1.68$ SD), Top Esports in 2024 ($-2.07$ SD), and Anyone's Legend in 2025 ($-1.16$ SD). Each was eliminated by T1, the most buff-exposed quarterfinalist in those years. T1 beat Weibo in the final and the other two in the quarterfinals. As with the T1 case, these pairings do not show what would have happened under another patch.

The identity of the most exposed team depends on the question asked. Among quarterfinalists, lowest net exposure and highest nerf exposure identify the same team in eight of ten years. Across all participating teams, the team with the lowest net exposure exited before the quarterfinals in five of ten years and never won. Together, the two perspectives make clear why scrutiny should cover the whole field and both directions of change.

\begin{table}[htbp]
\centering
\small
\caption{The quarterfinalist with the most negative net exposure each year.}
\label{tab:burden}
\begin{tabular}{llrrl}
\toprule
Year & Lowest-net-exposure quarterfinalist & Net (SD) & Versions & Finish \\
\midrule
2016 & Cloud9 & $-1.78$ & 3 & QF \\
2017 & Samsung Galaxy & $-1.19$ & 3 & \textbf{Winner} \\
2018 & KT Rolster & $-1.88$ & 4 & QF \\
2019 & Invictus Gaming & $-1.80$ & 4 & SF \\
2020 & G2 Esports & $-1.73$ & 3 & SF \\
2021 & Cloud9 & $-1.42$ & 4 & QF \\
2022 & Rogue & $-1.07$ & 3 & QF \\
2023 & Weibo Gaming & $-1.68$ & 6 & Runner-up \\
2024 & Top Esports & $-2.07$ & 3 & QF \\
2025 & Anyone's Legend & $-1.16$ & 3 & QF \\
\bottomrule
\end{tabular}
\end{table}

\FloatBarrier
\section{Discussion}
\label{sec:discussion}

\paragraph{Late rulemaking is a governance decision.} The authority to alter champions shortly before Worlds is authority over the resources with which teams compete. Our evidence locates an observable distribution of those changes. Teams more reliant on widely contested champions have more negative net exposure, and greater buff exposure is associated with additional bans on buffed champions, although the size of this association depends on how pick share is modeled. These patterns give concrete content to the institutional concern that the game owner also stages the championship. A patch can pursue legitimate balance goals while imposing unequal adjustment demands. The publisher should explain how it weighs competitive stability, adaptation, champion diversity, and the viewing experience when exercising that authority.

\paragraph{Accountability should not require a lost-title counterfactual.} A team eliminated after late changes cannot show from its elimination alone that it would have advanced without those changes. Likewise, a winning team's high buff exposure does not show that the publisher favored it. Our outcome analysis does not settle such disputes in either direction. Even after pooling ten events, the association between exposure and winning remains uncertain. Under our simulation model, per-game associations smaller than about $3.4$\pp{} in either direction are detected less than $80\%$ of the time, which corresponds to roughly $6.3$\pp{} in an evenly matched best-of-five under the illustrative conversion. More fundamentally, a dispute about a particular title concerns a single event without an observable counterfactual. Larger samples can sharpen estimates of average effects, but not the answer for one championship. This is one reason to set procedural standards that do not depend on proving that a patch changed who won. The legitimacy of a decision process can be examined through its timing, stated objectives, evidence, and response to objections before a disputed result occurs. That examination should also include the teams with the most adverse exposure, which draw little attention once they are eliminated.

\paragraph{From publishing changes to explaining decisions.} Riot's patch notes already discuss professional-play goals \citep{riot1319,riot1418}, and its published criteria state when professional presence calls for a nerf and how buffs serve diversity targets \citep{riot2020proplay,riot2020framework}. That is a meaningful step, and our results show that measured net exposure is partly predictable from team pools. Predictability, however, is not accountability. A team specialized in the meta can expect its champions to be targets, but not which changes will arrive or whether they will come before or after its last official matches. The public materials we reviewed document no independent route for contesting a trade-off made against it. A governance record should go further. It should identify the intended competitive problem, the alternatives considered, the anticipated team-level incidence, and the reasons for proceeding so close to the event. Our exposure measure offers one reproducible input to such a record. It is feasible for the publisher, which knows the changes it is planning, while teams' recent picks are public. Estimates are available while changes are being designed. Once teams have played their last official games, the exposure measures we propose here can be computed exactly before the tournament begins, without any forecast. A useful disclosure would cover all qualified teams, distinguish buffs from nerfs, record the patch versions each team has played, and show sensitivity to coding and pool definitions. It should also acknowledge item and system changes omitted here. Such a record would allow outside analysts and participants to question the decision on shared evidence. Disclosure alone provides no authority to seek reconsideration.

\paragraph{Procedural safeguards before competition.} We propose four linked safeguards.
\begin{enumerate}
\item \textbf{An earlier commitment and a defined freeze.} Riot already aims to minimize large and systemic changes in the lead-up to Worlds \citep{riot2020proplay}. A defined freeze would make that commitment checkable. Announce the championship patch timetable in advance and allow a meaningful preparation interval after substantive changes are fixed. State narrow exceptions for critical bugs or competitive exploits, with published reasons for using them. Our analysis does not identify an optimal number of weeks.
\item \textbf{A team-level impact statement.} Before the patch is locked, publish its goals, the trade-offs among them, and team-level exposure of the kind described above, with its sensitivity to coding. Update the statement as teams qualify and changes are revised. Distinguish estimates made while teams' pools and the set of qualified teams could still change from exact figures computed after teams' last official games.
\item \textbf{Independent review with participant representation.} Give a panel with player and team representation access to the proposed changes and their rationale, with independence from both event promotion and the patch's authors. Publish its recommendation and the publisher's reasoned response. The panel should assess consistency with declared criteria, with conflicts of interest disclosed.
\item \textbf{A time-bounded route to challenge.} Specify who can object, what evidence is required, who reviews the objection, and when a decision must be issued before Worlds. Permit concerns about procedure or concentrated exposure. Retain decisions and responses for retrospective review.
\end{enumerate}

These safeguards are proposals motivated by the concentration of authority and the observed incidence. Earlier freezes can lock in an imbalance that a later patch would have corrected, and teams have incentives to defend their own preferred champions. Review therefore needs common criteria, not a veto for any adversely exposed team. Equal numerical exposure is not the policy objective, since balancing dominant champions and rewarding adaptation can be legitimate. Instead, the objective is a timely, reviewable explanation of why these changes, with these consequences, belong immediately before the championship. We do not catalog every existing consultation channel. The proposed standard concerns what participants and outsiders should be able to inspect and contest.

\section{Limitations}
\paragraph{Inference.} Only ten events support the event-bootstrap intervals, and teams recur across them. Ratings and exposure are measured with uncertainty that the conditional power simulation does not propagate. Serial dependence, model misspecification, and observational confounding can also affect inference. The associations do not separate pool style from the length of a team's patch window. Champion comparisons and secondary models are exploratory, their $p$-values are not corrected for the number of analyses, and the cases compare teams selected post hoc from quarterfinalists.

\paragraph{Measurement.} Exposure counts direct champion changes only, and item, rune, and system changes can reinforce or offset them. Future work could extend exposure to items and runes. Leaguepedia scoreboards record each player's final items and runes in professional games. Each team's pre-Worlds builds could then weight item and rune changes, as pick shares weight champion changes here. Changes in purchase rates after each patch could check the coded directions, as changes in pick-and-ban presence do for champions. Human validation rests on one coder, the author, and on first-pass codes. The game-data check uses a heuristic direction rule and cannot see changes the game data do not record. Magnitude codes are less reliable than directions, and agreement does not validate their assumed spacing, although the main results do not depend on magnitudes. Pools observed before Worlds may already reflect adaptation to announced patches. The 2025 eligibility correction accounts for prior picks under fearless draft, but does not model every strategic consequence of that format. Excess bans and pick advantage are jointly determined by both teams, and the size of the ban association depends on how pick share is modeled. The association between meta concentration and buff exposure depends on how presence is defined and on allowing for dependence within a Worlds, whereas its associations with nerf exposure and net exposure do not.

\paragraph{Data.} The $547$ games restored from Leaguepedia cannot themselves be checked against Oracle's Elixir. The two sources agree closely on the games both hold, and games that neither source records are not recovered. The record corrections follow fixed rules, but the sources behind them are not fully independent, because Leaguepedia and the statistics services draw partly on the same official match data. Games whose winners the two sources record differently are excluded although they were actually played, and one of them falls within a participant's pre-Worlds window (Appendix~\ref{app:supplement}).

\paragraph{Scope.} The balancing criteria we cite were published in 2020, and we did not assess whether individual patches followed them. We cannot infer publisher intent, causal title effects, or perceived fairness, and we have not tested whether the proposed safeguards would work.

\section{Conclusion}
Even at its premier championship, LoL combines late changes to the game with publisher control over the event. Across ten Worlds tournaments, those changes intersected unevenly with team pools and were associated with differences in opponent bans. Net exposure was more negative for teams whose pools concentrated on the contested meta, including regional winners. Looking at both the most buff-exposed and the most negatively exposed quarterfinalists makes the distribution visible beyond the title winner. Our analysis of match results does not resolve whether that distribution changed who won. Accountability should therefore begin before play, with an earlier commitment, explicit assessment of team-level incidence, independent scrutiny, and a usable process for challenging decisions. The warning is about concentrated discretion over competitive conditions. The remedy is to make its exercise visible, justified, and contestable.

\section*{Ethics statement}
This study uses publicly available records of professional matches and publicly released patch notes. It involves no private data and no interaction with players, teams, or the publisher. We name teams, including T1 and the quarterfinalists in Table~\ref{tab:burden}, because the governance questions concern specific, publicly visible championships. Naming them is not intended to diminish any team's achievements. Championships are won by players and staff who adapt to whatever version of the game is in place, and our analysis provides no evidence that any title was undeserved or that the publisher favored any team. T1 features prominently because it was the most buff-exposed quarterfinalist in four consecutive years. We use it as an illustrative case to show why late rule changes deserve procedural scrutiny, with the aim of contributing to a fairer path for esports, not of questioning particular results. Likewise, the quarterfinalists with the most negative net exposure are included to show adverse exposure that is easily overlooked, not to explain or excuse their results.

\section*{Data and code availability}
Match data are available from Oracle's Elixir \citep{oracleselixir}. The $547$ games restored from Leaguepedia \citep{leaguepedia} are included in the repository under Leaguepedia's CC BY-SA 3.0 license. The coded patch notes, validation data, analysis code, and all derived tables are available in a public repository at \url{https://github.com/haewoon/lol-worlds-patch-accountability}.

\section*{Use of AI tools}
The research idea originated with the author. AI assistants (Claude from Anthropic and Codex from OpenAI) helped with coding and writing. They wrote and reviewed analysis code and revised parts of the manuscript. Patch notes were also coded by LLM agents, as described in Section~\ref{sec:coding} and validated in Appendix~\ref{app:coding}. The author checked all code, analyses, and text, and takes full responsibility for the content of this paper.

\bibliographystyle{plainnat}
\bibliography{refs}

\appendix
\FloatBarrier
\section{Games restored from Leaguepedia}
\label{app:supplement}
Oracle's Elixir does not record some games that Worlds participants played before the event. We compared it with Leaguepedia's scoreboard tables \citep{leaguepedia} over the $365$ days before each Worlds, the window of the pre-event ratings. Matching proceeds in two stages. First, a Leaguepedia game is matched to an Oracle's Elixir game when the same ten champions were picked within $36$ hours. Matches are one-to-one and pair the same game of a series first, because teams occasionally repeat a draft in the next game, and then the nearest in time. Leaguepedia team names are linked to Oracle's Elixir team identifiers from these matches only, and only when at least three matches and $90\%$ of the name's matches point to one identifier, within the year or otherwise across years. Other names keep a separate identifier. The links recover every participant of every Worlds. Second, an unmatched game may match a remaining Oracle's Elixir game of the same two linked teams within $36$ hours when at least eight of the ten picks agree, which allows for a pick recorded differently. An unmatched game is restored when a participant of that year's Worlds played in it and it belongs to a competition of the kinds Oracle's Elixir records. These are regional leagues with their playoffs, qualifiers, and cups, domestic and third-party cups such as the Demacia Cup, KeSPA Cup, NEST, and IEM, and international events, including Rift Rivals and the 2016 International Wildcard Qualifier, which Oracle's Elixir does not cover at all. National-team events, showmatches, and one-off invitationals are not restored. Where Oracle's Elixir holds games of the same two teams within $36$ hours, only the surplus is restored, because the same game may carry a different pick record. Restored games of such a partly recorded series take the series' Oracle's Elixir patch. Opponents that appear only in restored cup games take the cup as their league in the rating model, as teams that appear only in the cups Oracle's Elixir records already do.

This adds $547$ games: $185$ from the LPL, $147$ from Rift Rivals, $116$ from other regional leagues and their cups, $75$ from domestic and third-party cups (including Brazil's Superliga ABCDE), $23$ from the 2016 International Wildcard Qualifier, and one from MSI. Of these, $488$ are from 2016--2019. On the $15{,}793$ participants' games matched in both sources, Leaguepedia's first-listed team is the blue side in all of them, and the sources agree on the winner in $99.99\%$ and on the set of bans in $99.7\%$. Of the $15{,}518$ games for which both record a patch, they agree on it in $99.8\%$, and the $32$ differences are each one version apart. These rates describe the games both sources hold. The restored games themselves cannot be checked against Oracle's Elixir. Restoration changes the last officially played version of seven team-years: INTZ and Albus NoX Luna in 2016, EDward Gaming, Royal Never Give Up, and Team WE in 2017, Ascension Gaming in 2018, and MEGA in 2019. It also changes the $100$-day pools of $55$ team-years, and four team-years that had fewer than ten games in Oracle's Elixir now have eligible pools. The restored games are in the replication files with their Leaguepedia identifiers, under Leaguepedia's CC BY-SA 3.0 license.

Oracle's Elixir itself contains records that fail internal checks. Table~\ref{tab:corrections} lists the problems, the rule applied to each, and the number of records affected, and the replication files list every record with its evidence. In total, $93$ records are removed, $39$ winners set, $17$ games excluded, and $5$ records moved. A record that shares statistics across drafts can also appear in the copy and winner rows of the table. Two records hold one game when they have identical player rows (players, champions, and results), game length, and start time. They also hold one game when they share the draft, game length, and each player's kills, deaths, assists, gold, and result, even if player names or a few seconds of start time differ. A shared draft alone does not make two records a copy. At most one record is retained per copy group. If the records name different team pairs and Leaguepedia confirms none, the whole group is removed. Removing copies also allows restoration of displaced games, such as game 2 of EDward Gaming's 2018 Demacia Cup series against Shadow Cream.

For results, an Oracle's Elixir record counts as one source, whether we read its result or its statistics. Its statistics name the side with at least two more towers destroyed and more gold. A missing winner is set only when at least two sources agree and none contradicts. The other sources are the Leaguepedia record, a Games of Legends game page, and a Liquipedia series score that forces the result given the other games of the series. We do not use Liquipedia's order of maps, because it conflicts with both other sources in several 2018 LDL series. Games whose recorded winner differs from Leaguepedia's are therefore excluded, and statistics copied from another game are not used. As a consistency check, the statistics rule names a side in $85{,}721$ of the $90{,}458$ Oracle's Elixir games of 2016--2025 with one recorded winner and agrees with that winner in $99.71\%$. This is agreement within Oracle's Elixir, not accuracy against another record. Two of the relabeled records are 2023 Coupe de France games of Team BDS Academy recorded as Team BDS. No correction touches a Worlds game.

Two games that participants played are missing from the data. The first is game 1 of the 2016 Demacia Cup final, won by EDward Gaming according to Leaguepedia and Liquipedia's $3$--$0$ series score but by I May according to Oracle's Elixir. The second is game 3 of Snake Esports' $2$--$1$ win over Invictus Gaming in the 2017 Demacia Championship, which Oracle's Elixir records as a copy of game 2 and neither source's scoreboard holds.

\begin{table}[h]
\centering
\small
\caption{Corrections to Oracle's Elixir records.}
\label{tab:corrections}
\begin{tabular}{>{\raggedright\arraybackslash}p{4.8cm}>{\raggedright\arraybackslash}p{6.9cm}>{\raggedright\arraybackslash}p{2.7cm}}
\toprule
Problem & Rule & Result \\
\midrule
Copies naming the same teams & keep one record & 41 removed \\
Copies carrying another game's team rows & keep the record whose team rows fit its players & 3 removed \\
Copies naming different team pairs (18 groups, from Benelux, Northern European, and Taiwanese competitions) & keep the record whose pair Leaguepedia confirms & 3 kept, 49 removed \\
Statistics shared across drafts (5 groups, 22 records) & keep only when Leaguepedia records the same winner & 6 kept \\
Winner differs from Leaguepedia (13 games) & sources contradict & 13 excluded \\
No winner recorded (42 games, 35 of them from the 2018 LDL finals) & set when two sources agree and none contradicts & 39 set, 3 excluded \\
Team rows from another game (5 records with a recorded winner) & keep without bans when Leaguepedia, the result, and the statistics agree & 4 kept, 1 excluded \\
Wrong team label & move the side to the team Leaguepedia names when at least three of its five players are that team's regulars and at most one is a regular of the recorded team (regulars: players in at least 20\% of the team's other games within 180 days before or after the game) & 5 moved \\
\bottomrule
\end{tabular}
\end{table}

A scan of the $16{,}531$ games that participants played in the $365$ days before Worlds finds $54$ in which a participant used at most one of its regular players ($57$ participant--game records). For the $33$ games ($34$ records) from Oracle's Elixir, the team agrees with the matched Leaguepedia game. The other $21$ ($23$ records) were restored from Leaguepedia and keep its labels. They are earlier-season rosters, substitutes, and academy rosters that an organization entered under its own name, for example in the KeSPA Cup and the Demacia Cup, and they count as the team's games. The scan flags unusual lineups but does not establish that no other team-label error remains.

\section{Patch-note coding}
\label{app:coding}
\paragraph{Procedure.} LLM agents coded the patch notes on September 23, 2026 (Claude Opus 5.5, model identifier \texttt{claude-opus-5-5}, Anthropic), run as independent sub-agents in Claude Code. Five agents each coded the patches of two consecutive tournament years, from 2016--17 to 2024--25. A sixth agent independently re-coded 2019, 2022, and 2025 for reliability. Each agent retrieved the notes itself, using official Riot patch notes, through the Internet Archive for older patches, and the League of Legends Wiki patch pages for hotfixes and exact numbers. Agents included hotfixes released before the start of that year's Worlds. The instructions began: ``Work from the patch-note TEXT only (blind coding): do not use or look up esports results, which teams won, or what pro players picked.'' Agents were told not to open any other file in the project. The full prompts are in the replication files.

\paragraph{Coding scheme.} For every champion changed in a patch, the agent combined all lines into one net judgment. It recorded \texttt{direction} as buff (net power increase), nerf (net power decrease), adjust (mixed or sidegrade), rework (visual or gameplay update with unclear net effect), or bugfix. For buffs and nerfs it recorded \texttt{magnitude} as $1$ for small changes, roughly one number and under $10\%$ of one ability or stat, $2$ for moderate ones, several numbers or a meaningful change to a core ability or base stat, and $3$ for large ones, a multi-part change clearly meant to move the champion's viability. It also recorded \texttt{pro\_targeted}, whether the notes say the change targets professional or high-level play, \texttt{hotfix}, and a summary of at most $20$ words. A later hotfix to the same champion within a patch is a separate row. Changes to modes other than Summoner's Rift were excluded. Three rows describing ARAM-only changes were removed afterward and are listed in the replication files. Table~\ref{tab:coding} summarizes the codes.

\paragraph{Audit.} We audited every tool call of the six agents. Their web searches all concerned locating patch notes or hotfixes (e.g., ``League of Legends patch 12.18 hotfix September 2022 mid-patch''), and every page they fetched was on \texttt{leagueoflegends.com}, the League of Legends Wiki, or \texttt{web.archive.org}. No esports-results site was accessed. No agent opened the match data, any analysis output, or another agent's coding file. Because a language model may carry knowledge of esports history from pre-training, blindness here is procedural, not absolute. The game-data check below does not depend on any language model.

\begin{table}[h]
\centering
\small
\caption{Coded Summoner's Rift champion changes, 2016--2025 ($40$ patches).}
\label{tab:coding}
\begin{tabular}{lr}
\toprule
Direction & Changes \\
\midrule
Buff & 358 \\
Nerf & 262 \\
Adjustment & 54 \\
Rework & 10 \\
Bug fix & 406 \\
\midrule
Total & 1{,}090 \\
\bottomrule
\end{tabular}
\end{table}

\paragraph{Human validation.} The author coded a stratified sample of $120$ changes, $12$ per year, with the same scheme and without access to the LLM codes. A coding page showed the scheme's definitions and listed, for each change, the numeric lines extracted from the official notes, the release date, whether the row was a hotfix, and a link to the full notes. Direction agreed in $105$ of the $120$ changes ($87.5\%$) and the sign (buff, nerf, or neither) in $107$ ($89.2\%$, $\kappa = 0.84$). Signed scores correlated at $0.84$. Among the $64$ changes that both coders coded as the same buff or nerf, magnitude agreed exactly in $75\%$ and within one level in all. Of the $15$ disagreements, $11$ had an adjustment code on one side, typically for changes that mix increases and decreases, and $2$ were between a buff and a rework or bug fix. The one sign reversal arose where the coding page omitted the notes' explanatory text, and the author coded a stronger ally shield for Hwei in patch 14.18 as a nerf. For one hotfix row (Orianna, 7.16), the coding page also displayed the main-patch changes. We report these first-pass codes without adjudication.

\paragraph{Reliability.} A separately run LLM coder using the same model re-coded 2019, 2022, and 2025, a total of $309$ champion--patch pairs. Sign agreement was $99.4\%$ ($\kappa = 0.99$), agreement among pairs where either coder recorded a buff or nerf $98.9\%$, and the correlation of signed scores $0.97$. Team net exposures computed from the two codings correlate at $0.99$.

\paragraph{Game-data check.} To check direction independently of the notes' wording, we compared each coded champion's game values in the patch with those in the preceding patch. For 2016--2018 we used Riot's official Data Dragon \citep{ddragon}, which hold base statistics, ability cooldowns, costs, and ranges, and unnamed ability values. For 2019--2025 we used the game data extracted by CommunityDragon \citep{cdragon}, which hold base statistics and named ability values and coefficients. We excluded hotfix rows ($56$) and patches 10.17 and 10.18 ($40$ rows), since neither source holds 10.17. Because the snapshots reflect each patch at release, a mid-patch hotfix to the preceding patch first appears in the next snapshot. We therefore also excluded $17$ changes to champions hotfixed on Summoner's Rift during the preceding patch, listed in the replication files. Only Summoner's Rift values were compared, with values of other modes identified by name markers in the game data. This left $977$ changes. A changed value counted as stronger if it grew in magnitude, except cooldowns, costs, delays, cast times, recharge times, and required counts, which count as stronger when they fall. Tooltip-only values were ignored. Rank-dependent values were compared over the ability's recorded number of ranks, or ranks $1$--$5$ where the game data do not record it. Values whose rank-by-rank changes differed in sign were treated as reshaped. This rule is a heuristic. It counts changed values, not their size, and cannot see changes the game data do not record, such as passive abilities in Data Dragon. Table~\ref{tab:gamedata} cross-tabulates the codes against the direction of the numbers for the $977$ comparable non-hotfix changes. Its columns ``Only up'' and ``Only down'' refer to directional values, which may coexist with reshaped values, and ``Reshaped'' contains no directional change. We call a change one-directional when every value with a directional change moved the same way, and $79$ of the $450$ such changes also contain reshaped values. Counting any other code as a mismatch, the coded sign matched the one-directional data in $94.9\%$ of the $450$ changes ($91.1\%$ with Data Dragon, $96.2\%$ with CommunityDragon) and in $97.8\%$ of the $371$ without reshaped values. Most mismatches are changes coded as adjustments, reworks, or bug fixes. Among the $434$ one-directional changes coded as buffs or nerfs, the sign matched in $98.4\%$. Of changes coded as buffs, $74\%$ showed only increases and $1.8\%$ only decreases. Of nerfs, $76\%$ showed only decreases and $0.4\%$ only increases. Most of these seven contradictions, six buffs and one nerf, reflect the limits of the game data or the rule, for example a passive buff that Data Dragon does not record alongside a recorded ability nerf. Of changes coded as bug fixes, $95\%$ showed no numeric change. In the author's sample, $110$ changes were comparable and $55$ were one-directional. On these same changes the coded sign matched in $96.4\%$ for the LLM codes and $90.9\%$ for the author's codes, and in $100\%$ and $97.8\%$ on the $46$ without reshaped values.

\begin{table}[h]
\centering
\small
\caption{LLM codes against game-data directions between consecutive patches ($977$ changes).}
\label{tab:gamedata}
\begin{tabular}{lrrrrr}
\toprule
 & \multicolumn{5}{c}{Game values} \\
\cmidrule(lr){2-6}
LLM code & Only up & Only down & Mixed & Reshaped & No change \\
\midrule
Buff & 245 & 6 & 25 & 18 & 39 \\
Nerf & 1 & 182 & 30 & 8 & 17 \\
Adjustment & 4 & 5 & 32 & 2 & 8 \\
Rework & 1 & 0 & 8 & 0 & 0 \\
Bug fix & 5 & 1 & 0 & 12 & 328 \\
\bottomrule
\end{tabular}
\end{table}

\paragraph{Coding sensitivity.} We reran the entire analysis pipeline, from pre-tournament ratings through the ban models, power simulations, and permutation calibration, under three alternative codings (Table~\ref{tab:codingsens}, where bracketed values are $95\%$ event-bootstrap intervals). In (A), the $23$ changes whose game values moved in only one direction but were not coded with that sign take the direction of the numbers. A reversed sign keeps its magnitude, and a formerly unsigned code receives magnitude $1$. In (B), every comparable change takes its direction from the numbers, and changes whose values are mixed, reshaped, or unchanged score zero. Signed changes keep the LLM magnitude, or $1$ if newly signed. Changes without comparable game data, namely hotfixes, patches 10.17--10.18, and champions hotfixed during the preceding patch, keep their LLM codes. B is therefore a stress test of the coded directions, not an LLM-free coding. Because it discards changes the game data do not record, such as passive changes before 2019, it is not a correction either. In (C), magnitudes are ignored, and each buff scores $+1$ and each nerf $-1$. Across the three codings, nerfs remain concentrated on the contested meta, the ban interaction for buffed champions remains, the outcome interval still includes zero, the null rejection rate stays near $5\%$, and the $80\%$ detection threshold stays near $3.4$\pp{} per game. The most buff-exposed quarterfinalists still reach six finals and win four titles, with T1 in each year from 2022 to 2025. The most negatively exposed quarterfinalists reach two finals and win one title under A and C, as in the main coding. Under B, the 2017 team is Longzhu Gaming instead of Samsung Galaxy, leaving one final and no title. Under B, the identity of these teams changes in three and two of the ten years, respectively. Variant B weakens the meta slope and the share of top-decile champions nerfed, because Riot's changes to contested champions often mix increases and decreases and B scores them as zero. The buff coefficient in the model with separate buff and nerf exposure stays at $+2.2$\pp{} under A but falls to $+0.7$\pp{} under B. B removes many buffs that even the official game data cannot see. Of the $88$ buffs that B removes or reverses, $39$ show no change at all in the game data. Before 2019, for example, Riot's Data Dragon did not record passive abilities, and Jarvan IV's shorter passive cooldown in patch 8.19 counts as no change. The buff coefficient is therefore sensitive to coding and should not be read as evidence.

\begin{table}[h]
\centering
\small
\caption{Main results under three alternative codings.}
\label{tab:codingsens}
\setlength{\tabcolsep}{4pt}
\begin{tabular}{lrrrr}
\toprule
 & Main & A & B & C \\
\midrule
Changes recoded (of 1{,}090) & -- & 23 & 160 & 369 \\
Team net exposure, $r$ with main & 1 & 0.99 & 0.92 & 0.96 \\
Pre-Worlds picks nerfed / buffed (\%) & 33.4 / 7.5 & 33.3 / 7.8 & 28.3 / 6.1 & 31.9 / 7.5 \\
Top-decile champions nerfed / buffed (\%) & 52 / 1 & 52 / 2 & 45 / 2 & 50 / 1 \\
Meta concentration $\rightarrow$ net exposure (SD) & $-0.40$ & $-0.39$ & $-0.30$ & $-0.41$ \\
Regional winners: median range (SD) & 1.80 & 1.79 & 1.99 & 1.48 \\
Regional winners: meta--exposure $r$ & $-0.50$ & $-0.46$ & $-0.41$ & $-0.49$ \\
Ban interaction, buffed (pp; $p$) & $+11.8$ (0.004) & $+11.8$ (0.003) & $+9.7$ (0.003) & $+11.3$ (0.007) \\
Ban interaction, other (pp; $p$) & $-2.5$ (0.055) & $-2.5$ (0.056) & $-2.1$ (0.129) & $-2.3$ (0.087) \\
Net exposure $\rightarrow$ win (pp per SD) & $+1.3$ & $+1.3$ & $+1.4$ & $+1.3$ \\
\quad interval & [$-0.9$, $+3.7$] & [$-0.9$, $+3.6$] & [$-0.4$, $+3.2$] & [$-1.0$, $+3.5$] \\
\quad permutation $p$ & 0.333 & 0.326 & 0.301 & 0.356 \\
\quad null rejection rate (\%) & 5.1 & 5.2 & 5.1 & 5.1 \\
\quad 80\% detection threshold (pp) & 3.4 & 3.4 & 3.4 & 3.4 \\
Buff exposure $\rightarrow$ win (pp; $p$) & $+2.2$ (0.111) & $+2.2$ (0.114) & $+0.7$ (0.587) & $+2.4$ (0.079) \\
Most buff-exposed QF team: finals / titles & 6 / 4 & 6 / 4 & 6 / 4 & 6 / 4 \\
Lowest-net QF team: finals / titles & 2 / 1 & 2 / 1 & 1 / 0 & 2 / 1 \\
\bottomrule
\end{tabular}
\end{table}

\section{Pre-tournament strength and robustness}
\label{app:strength}
Table~\ref{tab:rating} evaluates the rating model on Worlds games. Its two hyper-parameters are the half-life of the down-weighting ($120$, $240$, or $480$ days) and the penalty on team deviations ($1$, $2$, $4$, $8$, $16$, or $32$). The main rating uses the pair with the lowest log loss over all Worlds games, $240$ days and $8$, and the leave-one-year-out version chooses the pair for each year from the other years. Most of its accuracy comes from differences between regions, since league effects alone correctly classify $67.8\%$ of games. The rating is somewhat over-confident, with a calibration slope of $0.77$ ($95\%$ CI $0.65$--$0.90$). Our models estimate the rating's coefficient, which permits recalibration without eliminating rating error. In cross-region games LCK teams won $5.2$\pp{} more than predicted ($95\%$ CI $0.3$--$10.0$) and LPL teams $3.2$\pp{} less (CI $-7.6$ to $1.1$). The rank correlation between pre-tournament rating and final placement averages $0.70$ per year, and the eventual winner's rating rank averages $3.0$.

\begin{table}[h]
\centering
\small
\caption{Accuracy of pre-tournament predictions on $1{,}097$ Worlds games.}
\label{tab:rating}
\begin{tabular}{lcc}
\toprule
Model & Accuracy & Log loss \\
\midrule
Coin flip & 50.0\% & 0.693 \\
Blue side only & 54.0\% & 0.690 \\
League effects only & 67.8\% & 0.615 \\
Rating (hyper-parameters chosen on all years) & 68.7\% & 0.608 \\
Rating (hyper-parameters chosen leave-one-year-out) & 68.6\% & 0.611 \\
\bottomrule
\end{tabular}
\end{table}

\begin{figure}[h]
\centering
\includegraphics[width=\textwidth]{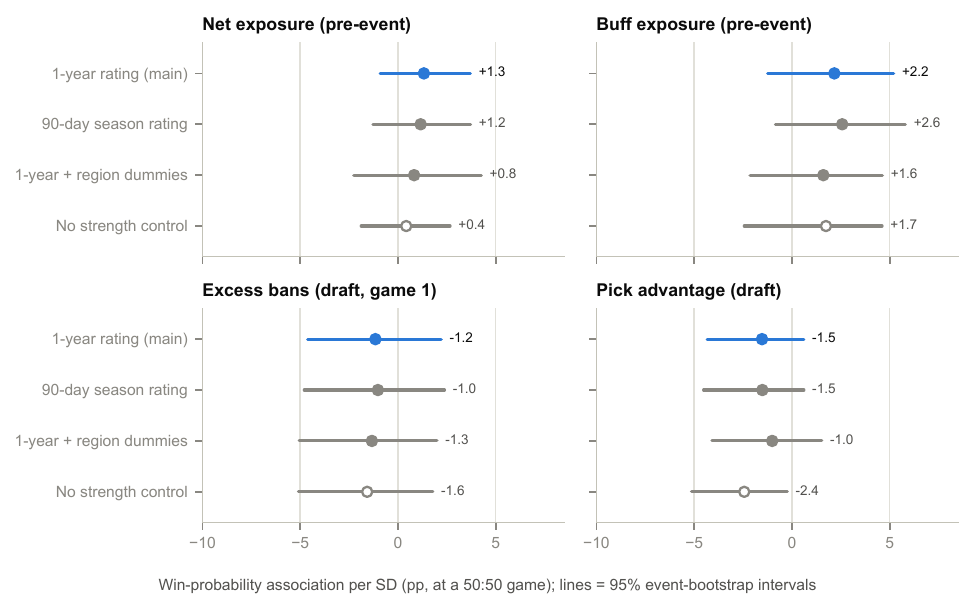}
\caption{Outcome associations under alternative pre-tournament strength controls.}
\label{fig:strength}
\end{figure}

Figure~\ref{fig:strength} re-estimates the four outcome associations with three alternative strength controls. The first is a rating that re-estimates team strength from the last $90$ days only and keeps the one-year league effects, because the $90$ days before Worlds contain few or no international games. The second adds region dummies to the one-year rating, and the third uses no control. A rating based on the last $90$ days alone, league effects included, has worse log loss than a coin flip ($0.749$ versus $0.693$) and is omitted from the figure. In the figure, excess bans use the first game of each series, and pick advantage is the difference in the two teams' picked-champion scores, each over that team's own unplayed versions. Buff exposure is modeled jointly with nerf exposure, and intervals resample events.

\section{Presence reference and dependence within a Worlds}
\label{app:presence}
Presence in the main analysis counts every game in the match data in the $100$ days before Worlds, each weighted equally. These include all leagues Oracle's Elixir covers, with development leagues such as the LDL and the LCK Challengers League, and the restored games of Appendix~\ref{app:supplement}. Riot's published criteria refer mainly to the major regions \citep{riot2020proplay}, and some of these games were played after a coded patch was released. Table~\ref{tab:presence} therefore repeats the champion-level and team-level comparisons with two other presence references. One uses the four major regional leagues only (LCK, LPL, LEC or EU LCS, and LCS or NA LCS, with the whole LTA in 2025). The other uses the $100$ days before each year's first coded patch was released. Team pools and breadth are unchanged, and meta concentration is recomputed from the same games. The release-by-release version measures presence in the $100$ days before each of $71$ coded releases, counting hotfixes separately, and asks whether that release nerfed or buffed the champion. None of these versions tests whether individual patches followed Riot's criteria. In Table~\ref{tab:presence}, percentages are shares of champion-years, or of champion-releases in the release-by-release column, in the top presence decile or the bottom eight deciles whose coded net change was a nerf or a buff. Coefficients are SD of exposure per SD of meta concentration, adjusted for breadth, with organization-clustered $p$-values in parentheses.

The team-level regressions cluster standard errors by organization. Because teams at the same Worlds also share its patches and meta, we report two further inferences for the main definition. One uses two-way clustering by organization and Worlds with a $t$ reference on nine degrees of freedom \citep{cameron2011}. The other is a wild cluster bootstrap over the ten Worlds with Webb weights, the null imposed, and $9{,}999$ draws \citep{cameron2008,webb2023}. For net, nerf, and buff exposure, the two-way $p$-values are $0.002$, $0.004$, and $0.048$, and the bootstrap $p$-values $0.006$, $0.009$, and $0.096$. With ten Worlds, neither is exact. The buff association thus weakens under Worlds-level inference, and it disappears when presence is measured before each year's first change (Table~\ref{tab:presence}). The nerf and net associations hold throughout.

\begin{table}[h]
\centering
\small
\caption{Champion-level and team-level comparisons under alternative presence definitions.}
\label{tab:presence}
\setlength{\tabcolsep}{4pt}
\begin{tabular}{lcccc}
\toprule
 & All games (main) & Major leagues & Before first change & By release \\
\midrule
Observations & 1{,}502 & 1{,}245 & 1{,}484 & 10{,}491 \\
Top decile: nerfed / buffed (\%) & 52 / 1 & 62 / 2 & 41 / 7 & 8.7 / 1.6 \\
Bottom eight: nerfed / buffed (\%) & 6 / 23 & 7 / 26 & 8 / 23 & 1.3 / 3.8 \\
Meta $\to$ net exposure & $-0.40$ ($<$0.001) & $-0.42$ ($<$0.001) & $-0.24$ (0.001) & -- \\
Meta $\to$ nerf exposure & $+0.35$ ($<$0.001) & $+0.35$ ($<$0.001) & $+0.24$ (0.002) & -- \\
Meta $\to$ buff exposure & $-0.20$ (0.004) & $-0.23$ (0.001) & $-0.01$ (0.923) & -- \\
\bottomrule
\end{tabular}
\end{table}

\section{Team lineages and persistence}
\label{app:orgs}
Table~\ref{tab:lineages} lists the ten lineages and the change Leaguepedia records for each \citep{leaguepedia}. Detailed evidence and page links are in the replication files.

\begin{table}[h]
\centering
\small
\caption{Team lineages linked across documented renames and whole-team acquisitions.}
\label{tab:lineages}
\begin{tabular}{lll}
\toprule
Lineage & Recorded change & Year \\
\midrule
SK Telecom T1 $\rightarrow$ T1 & Rebrand & 2019 \\
Samsung Galaxy $\rightarrow$ Gen.G & Roster and LCK slot acquired, later Gen.G & 2017 \\
Longzhu Gaming $\rightarrow$ DRX & Rebrand after a sale, as Kingzone DragonX & 2018 \\
ROX Tigers $\rightarrow$ Hanwha Life Esports & Rebrand, roster and LCK slot acquired & 2018 \\
I May $\rightarrow$ Bilibili Gaming & Roster and LPL slot acquired & 2017 \\
Suning $\rightarrow$ Weibo Gaming & Rebrand & 2021 \\
Splyce $\rightarrow$ MAD Lions $\rightarrow$ Movistar KOI & Rebrand keeping the LEC slot, later renames & 2019 \\
Counter Logic Gaming $\rightarrow$ NRG & Roster and LCS slot acquired & 2023 \\
Lyon Gaming $\rightarrow$ Movistar R7 & Forced rename to Rainbow7, later Movistar R7 & 2017 \\
Young Generation $\rightarrow$ Vikings Esports & Renames via Phong V\~{u} and Saigon Buffalo & 2018 \\
\bottomrule
\end{tabular}
\end{table}

With teams linked across documented renames and whole-team acquisitions as in the main analysis, we ask whether style and buff exposure persist within organizations. When teams renamed after a sponsorship or ownership change are linked, correlations across successive appearances are $0.28$ for breadth and $0.25$ for meta concentration, compared with $0.17$ for buff exposure ($p = 0.06$, $132$ pairs). Treating them as separate teams gives $0.35$, $0.26$, and $0.12$ ($p = 0.20$, $119$ pairs). Pool breadth and meta concentration persist under either definition, whereas buff exposure persists less clearly. Whether organizations differ in buff exposure also depends on what counts as one organization. An organization-variance test compares the variance of organization means, over organizations with at least three appearances, with its distribution when exposures are permuted within year. Treating renamed teams as separate, the test gives $p = 0.15$ before style adjustment and $p = 0.075$ after it. Linking documented renames gives $p = 0.01$ and $p = 0.006$. The linked result rests largely on one lineage. Longzhu Gaming (2017) and its renamed successor DRX (2020, 2022), whose 2017 roster shared no player with the later ones, averaged $+2.2$ SD of buff exposure, and without them $p = 0.26$. Without T1, $p = 0.006$. The test therefore establishes neither organization-specific differences nor their absence, and it cannot test intent.

\section{Role-level exposure}
\label{app:roles}
Role exposure applies the net-exposure formula within each position recorded in the match data, $\sum_c w_i(r,c)\,\delta_i(c)$ for role $r$, with the same pick shares and team-specific changes. The five role exposures sum to the team's net exposure, and a champion played in several roles counts in each role in which it was played. Table~\ref{tab:roles} reports the average share of each role's pre-tournament picks on champions that the team's unplayed patches nerfed or buffed, as means over $215$ team-years with $95\%$ intervals from resampling team-years. It also counts the team-years in which each role alone carried the largest absolute exposure. These counts cover $201$ team-years, excluding $13$ with tied largest roles and one with no exposure. Nerfed shares are highest in the bottom lane and jungle and lowest in top lane and support. Buffed shares are highest in top lane. Among the $214$ team-years with nonzero exposure, the role with the largest absolute exposure accounted for a median of $40\%$ of the sum of absolute role exposures (interquartile range $34$--$45\%$), compared with $20\%$ if exposure were spread evenly. Of these, $13$ had two or more roles tied for the largest. Under each alternative coding of Appendix~\ref{app:coding}, bottom lane, jungle, and mid remain the three roles with the highest nerfed shares, in that order. These are descriptive summaries. The intervals resample team-years and ignore dependence within events and organizations.

\begin{table}[h]
\centering
\small
\caption{Pre-tournament picks on nerfed or buffed champions, by role.}
\label{tab:roles}
\begin{tabular}{lrrr}
\toprule
Role & Nerfed picks (\%) & Buffed picks (\%) & Sole largest \\
\midrule
Top & 23.7 [20.7, 26.7] & 10.5 [9.1, 12.0] & 2 \\
Jungle & 40.6 [37.8, 43.2] & 5.7 [4.8, 6.7] & 71 \\
Mid & 34.5 [31.9, 37.2] & 7.8 [6.6, 9.0] & 48 \\
Bottom & 45.8 [43.2, 48.5] & 9.4 [8.0, 10.7] & 62 \\
Support & 22.3 [20.0, 24.6] & 4.1 [3.3, 5.0] & 18 \\
\bottomrule
\end{tabular}
\end{table}

\section{Ban model}
\label{app:bans}
The champion-level ban model of Section~\ref{sec:draft} has one observation for each Worlds game, focal team, and eligible champion. Its outcome is whether the opponent banned the champion. The baseline linear probability model includes the focal team's role-summed pre-event pick share, the opponent's pick share, and champion-year and banning-team-game fixed effects. The main interaction model adds pick share interacted with team-specific buffed and nerfed status, plus buff exposure interacted separately with buffed and other champion pick shares. Standard errors are clustered by focal team-year. Removing champion choices that were unavailable (champions disabled for all or part of an event and, in 2025, champions locked by earlier picks in the series) leaves $258{,}494$ observations.

In the baseline model, a unit increase in pick share is associated with $+43.9$\pp{} in ban probability. In the patch-status model, nerfed champions have a $-11.5$\pp{} pick-share interaction ($p = 0.007$). The main exposure-interaction model yields $+11.8$\pp{} for buffed champions ($95\%$ interval $+3.8$ to $+19.7$) and $-2.5$\pp{} for other champions ($-5.2$ to $+0.1$).

We apply the \emph{same exposure-interaction specification} to three restricted samples or outcomes. Excluding 2025 gives $+12.0$\pp{} for buffed champions ($p = 0.003$) and $-2.8$\pp{} for others ($p = 0.036$). Series game 1 gives $+13.1$ and $-2.3$\pp{} ($p = 0.012$ and $0.103$). First-phase bans give $+7.7$ and $-1.3$\pp{} ($p = 0.032$ and $0.266$). These checks support a more consistent positive association for buffed champions than a reduction for other champions, although the first-phase estimate is smaller. None identifies causal displacement.

The interaction multiplies a champion's pick share by the team's buff exposure, which itself contains that champion's pick share times its buff magnitude. The linear pick-share control does not separate this construction from team-wide exposure. We therefore report three post-hoc specifications. Recomputing exposure without the champion's own contribution (keeping the year's mean and SD) gives $+10.7$\pp{} for buffed champions ($p = 0.015$) and $-2.6$\pp{} for others ($p = 0.051$). Adding squared pick share separately for buffed and other champions gives $+9.6$\pp{} ($p = 0.031$; $95\%$ interval $+0.9$ to $+18.3$) and $-3.3$\pp{} ($p = 0.022$). Because team-specific windows make buffed status vary within champion-year, adding buffed and nerfed indicators and an interaction between buffed status and exposure gives $+13.3$\pp{} ($p = 0.002$) and $-3.0$\pp{} ($p = 0.024$). The sign of the buffed-champion interaction is stable, while its size ranges from $+9.6$ to $+13.3$\pp{} across specifications. The $p$-values are not corrected for these several checks.

Buff exposure had little team-level correlation with the share of Worlds picks on champions the team had played at least twice in the same role during the preceding $100$ days ($r = 0.03$). Within team-years, one additional excess ban on buffed champions was associated with a $1.8$-percentage-point lower familiar-pick share ($\beta = -0.018$, $p = 0.31$). These results do not establish equivalence in draft difficulty.

\section{Championship patch timing}
\label{app:timing}
Table~\ref{tab:timing} compares the release date we recorded for each designated Worlds patch with the event start in our manifest, including play-ins. Patch identifiers follow Oracle's Elixir, and 15.20 corresponds to Riot's 25.20 label in 2025. The interval ranges from $5$ to $21$ calendar days (median $9.5$). These dates describe release of the designated Worlds patch, not the first announcement, preview, access to a test server, or private notice to teams. They do not measure each team's preparation time, nor do they include the timing of later hotfixes. The exposure analyses use a different, team-specific window, which contains all coded patches newer than the team's last officially played patch through the Worlds patch. Date metadata and source URLs are retained in the replication files.

\begin{table}[htbp]
\centering
\small
\caption{Release of the designated Worlds patch and start of the event.}
\label{tab:timing}
\begin{tabular}{lcccc}
\toprule
Year & Patch & Recorded release & Worlds start & Days \\
\midrule
2016 & 6.18 & September 8 & September 29 & 21 \\
2017 & 7.18 & September 13 & September 23 & 10 \\
2018 & 8.19 & September 26 & October 1 & 5 \\
2019 & 9.19 & September 25 & October 2 & 7 \\
2020 & 10.19 & September 16 & September 25 & 9 \\
2021 & 11.19 & September 22 & October 5 & 13 \\
2022 & 12.18 & September 21 & September 29 & 8 \\
2023 & 13.19 & September 27 & October 10 & 13 \\
2024 & 14.18 & September 11 & September 25 & 14 \\
2025 & 15.20 & October 8 & October 14 & 6 \\
\bottomrule
\end{tabular}
\end{table}

\FloatBarrier
\section{Outcome-test calibration and conditional power}
\label{app:power}
We assess power under independent Bernoulli outcomes generated from the fitted rating-only model, the outcome model with the rating difference but no exposure, plus a known exposure coefficient, holding the observed match schedules, ratings, and exposures fixed. Each simulation refits the outcome model for the observed exposure and $1{,}000$ within-year permutations, using the same studentized statistic as the main test. In $2{,}000$ null simulations, the rejection rate was $5.1\%$ ($95\%$ Monte Carlo interval $4.2$--$6.2\%$). The unstudentized coefficient test rejected $6.0\%$. These rates support calibration under that model, not an exact test for observational data. With $500$ simulations per nonzero effect, power is $49\%$ at a positive $2.5$\pp{} slope and $83.2\%$ at $3.5$\pp{}. Interpolated $80\%$ thresholds are approximately $3.4$\pp{} in either direction \citep{bloom1995}. Pointwise Monte Carlo intervals at $+2.5$ and $+3.5$\pp{} are $44.6$--$53.4\%$ and $79.7$--$86.2\%$. An illustrative conversion of the positive threshold to a best-of-five series gives about $6.3$\pp{}, assuming identical independent games with constant win probability around $0.5$. This conversion is not an estimate of a team's title probability, and adaptive drafts can violate its assumptions.

\end{document}